# Scalable Optical Learning Operator

Uğur Teğin[1,2,*], Mustafa Yıldırım[2], İlker Oğuz[1,2], Christophe Moser[2] and Demetri Psaltis[1]

[1] Optics Laboratory, École Polytechnique Fédérale de Lausanne, Switzerland

[2] Laboratory of Applied Photonics Devices, École Polytechnique Fédérale de Lausanne, Switzerland

* ugur.tegin@epfl.ch

## Abstract

Today's heavy machine learning tasks are fueled by large datasets. Computing is performed with power hungry processors whose performance is ultimately limited by the data transfer to and from memory. Optics is one of the powerful means of communicating and processing information and there is intense current interest in optical information processing for realizing high-speed computations.  Here we present and experimentally demonstrate an optical computing framework (Scalable Optical Learning Operator) based on spatiotemporal effects in multimode fibers for a range of learning tasks from classifying COVID-19 X-ray lung images and speech recognition to predicting age from face images. The presented framework addresses the energy scaling problem of existing systems without compromising speed. We leveraged simultaneous, linear, and nonlinear interaction of spatial modes as a computation engine. We numerically and experimentally showed the ability of the method to execute several different tasks with accuracy comparable to a digital implementation.

## Introduction

Early optical computers were used to calculate linear operations such as the Fourier transform and correlations. They found applications in pattern recognition and synthetic aperture radar [1, 2]. However, with the advent of modern VLSI technology and efficient algorithms (e.g Fast Fourier Transform), digital signal processing based on silicon circuits became so fast and parallel that the analog optical computation that included the input and output electronic overhead became obsolete. Digital optical computing, that combined nonlinear optical switches [3] with linear optical interconnections [4] replacing wires, was then intensely pursued in the 1980's. Optical interconnections can be advantageous in terms of power consumption [5], however in an all-optical implementation this advantage is counter-balanced by the power inefficiency and large size of optical switches compared to the electronic ones. Therefore, all-optical digital computers are not yet competitive. Optics has also been used for the implementation of nonlinear computations that are not based on Boolean logic, such as the optical implementation of neural networks [6, 7]. In principle, the dense connectivity of neural networks and their relative robustness against noise and device imperfections, renders neural networks a promising area for optical computing.

Interest in optically implemented neural networks has intensified in recent years partially because the large size of databases that need to be managed stresses the capabilities of existing digital, electronic computers. Several promising approaches are being investigated and they are summarized in a recent review article [8]. The key challenge in designing a viable optical computer (including a neural one) is to combine the linear part of the system from where the competitive edge of optics derives, with nonlinear elements and input-output interfaces while maintaining the speed and power efficiency of the optical interconnections.

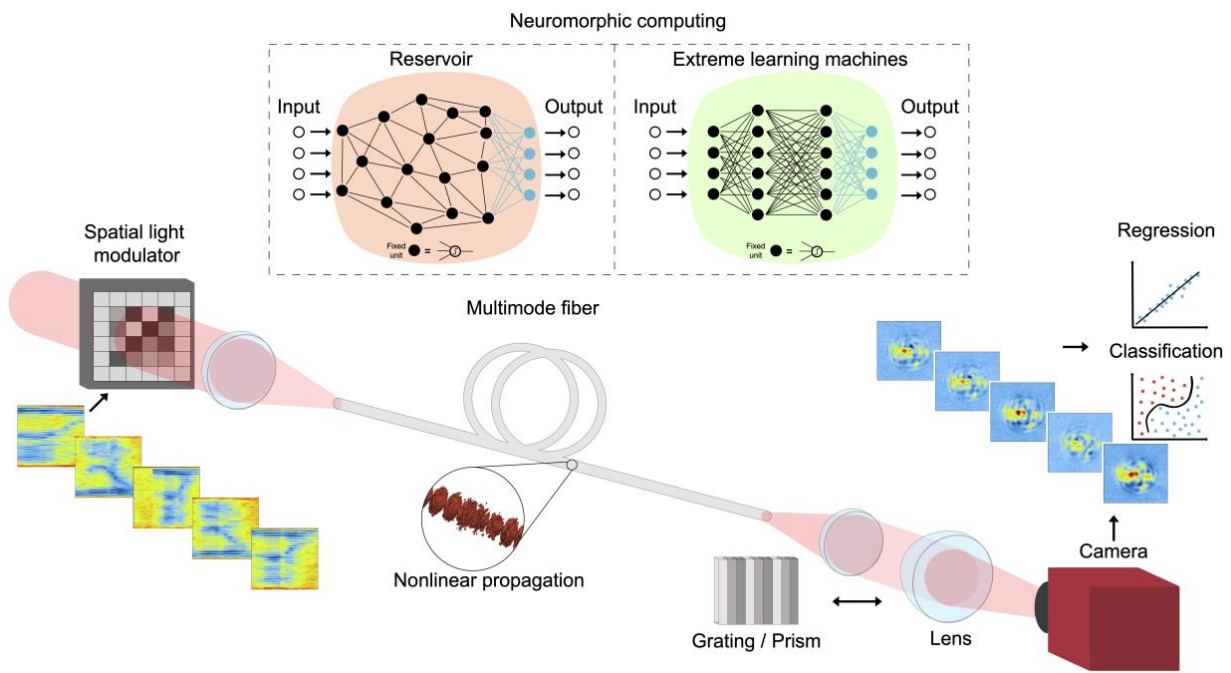

**Figure 1 Illustration of the fixed-parameter neural network architectures and the experimental setup for nonlinear projection with spatiotemporal multimode fiber nonlinearities.** The inset depicts neural network architectures with similar attributes as the MMF processor with black and blue connections indicating fixed and adaptable weights, respectively.

The solution we propose and demonstrate in this paper is the combination of the linear and nonlinear parts of the optical system in a shared volume confined in a multimode fiber (MMF). The principal advantage of this approach is the combination of the 3D connectivity of optics with the long interaction length and lateral confinement afforded by the fiber which makes it possible to realize optical nonlinearities at relatively low optical power. At the same time, the large number of spatial modes that can be densely supported in a MMF maintains the traditional high parallelism feature of optics, while maintaining a compact form factor. Finally, with the availability of megapixel spatial light modulators (SLMs) and cameras, the 2D input and output interfaces to the MMF can sustain a large information processing throughput. We refer to the proposed method as SOLO (Scalable Optical Learning Operator) in the remainder of this paper.

A schematic diagram of the MMF processing element is shown in Figure 1. The data to be processed is entered through the 2D spatial light modulator on the left. At sufficiently high illumination peak power, the light from a pulsed light source is nonlinearly transformed as it propagates through the fiber and the result of the computation is projected on the 2D camera. Given the properties of the fiber and the laser source, the input-output operation performed by the MMF is fixed and highly nonlinear. We implement a reconfigurable processor by combining the fixed nonlinear MMF mapping in the optical domain with a single layer digital neural network (decision layer) trained to recognize the output recorded on the camera using a large data set of input-output pairs. For instance, we used this system to diagnose with high accuracy (93%) COVID-19 from X-ray images of lungs. A large database of X-ray images of lungs with various diseases including COVID-19 was used to train the single layer network that classifies the representation of the lungs that is produced at the output camera. The notion of combining a complex, fixed mapping with a simpler programmable processor to realize a powerful overall system, including the optical implementation of such machines, has been used in support vector machines [9,10] reservoir computing [11-15], random mappings [16-19], and extreme learning machines [20,21]. The nonlinear mapping performed by the MMF is not the same as in any of the earlier approaches. As we will show, it proves to be very effective in transforming the input data space on the SLM to a nearly linearly separable output data space (camera at end of the MM fiber) at very high speed and power efficiency.

In the remainder of the paper, we present numerical and experimental results from our optical computing framework for single variable linear regression, multivariable linear regression, age prediction from face images, audio speech classification and COVID-19 diagnosis from X-ray images tasks. We then discuss how the system scales to large data size and estimate the power consumption per operation. These studies show that the analog optical computer based on the MMF is power efficient, versatile and obtains accuracy performance comparable to that obtained with digital computers when solving the tasks we investigated.

## Results

**Experimental studies**

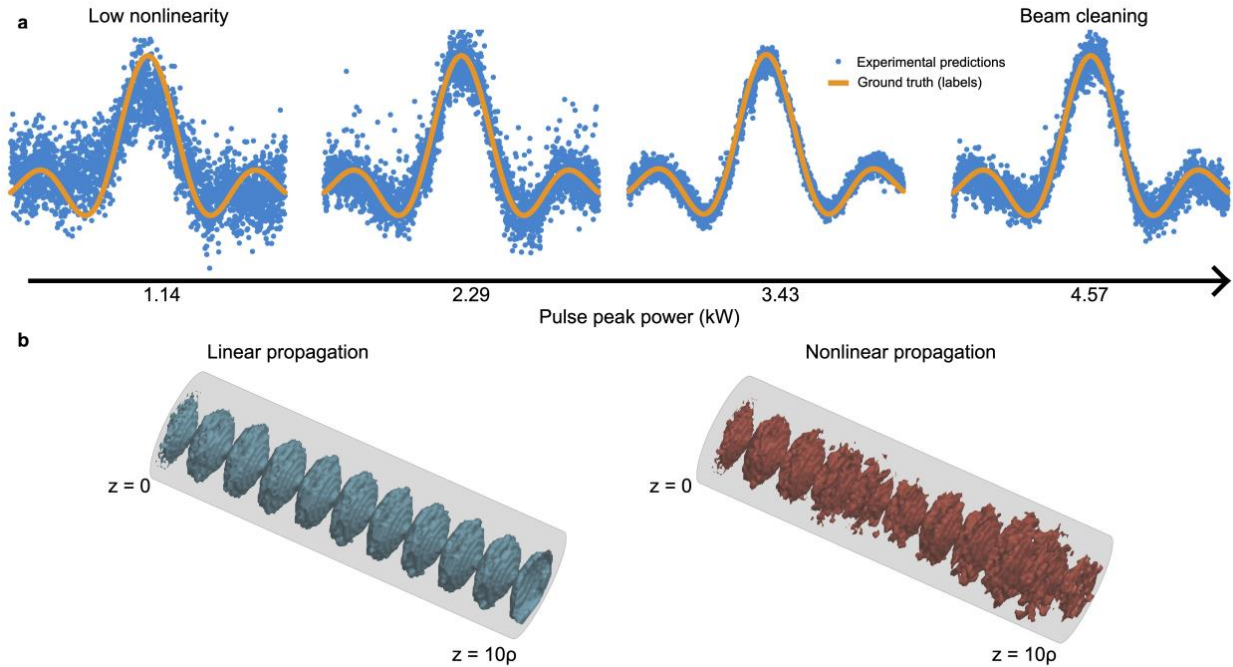

**Figure 2 Learning a nonlinear function (sinc):** dependence on the pulse propagation regimes in a graded-index multimode fiber caused by increasing the input optical peak power. (a) Experimental measurements: increasing the input peak power increases the nonlinear coupling between modes which translates in better learning. Beyond an optimal peak power, the learning performance degrades due to the Raman beam cleaning effect (see text). (b) Illustration of the propagation difference for linear (low peak power) and nonlinear (high peak power) cases in a GRIN MMF with 10 self-imaging period length.

Multimode fibers (MMFs) exhibit waveguide properties while allowing a large number of spatial degrees of freedom. Graded-index multimode fibers (GRIN MMFs) in particular, have become the subject of significant interest for telecommunications, imaging and nonlinear optics studies due to their unique properties such as relatively low modal dispersion and periodic self-imaging. In recent years, with spatiotemporal pulse propagation in GRIN MMFs, various nonlinear frequency generation dynamics [22-26], nonlinear beam cleaning [27] and spatiotemporal mode-locking [28-30] have been realized. Moreover, learning and controlling nonlinear optical dynamics in GRIN MMFs was demonstrated by modifying the spatial properties of the intense pump pulse with a spatial light modulator (SLM) or deformable mirror device (DMD) [31,32].

In machine learning studies, a variety of nonlinear transformations of the input data have been investigated in order to enable learning of complex relations hidden in the data [33]. In our case, we make use of the nonlinear mapping that takes place at high light intensities when an input pattern propagates in a multimode fiber as a physical realization of machine learning. The experimental setup in Fig.1 is explained in detail in the Methods section. In this setup, information spatially modulated an intense laser pulse with the input data and the Fourier transform of the spatially modulated beam was focused on the input facet of the optical fiber through a lens. The amount of light coupled to each of the modes of the fiber is given by the inner product between the incident light amplitude and the mode profile. Upon propagation, the initial complex modal coefficients evolve according to spatiotemporal linear and nonlinear effects. The nonlinear transformation of information is achieved by nonlinear energy exchange between the fiber modes. The transformed information at the end of the fiber is imaged onto a camera, and the image was downscaled such that the spatial sampling period is approximately equal to the resolution limit, which can be approximated by λ/2(NA), the Abbe diffraction limit. Each pixel of the downscaled image served as an input feature to a linear regression or equivalently, to a single layer neural classification algorithm to estimate the identity of the input on the SLM.

*Learning a nonlinear function*

To test this, we selected a simple regression problem on a dataset generated with a nonlinear (Sinc function) relation. The input information (x) were randomly generated numbers between -π to π and the corresponding output labels (y) were generated according to the y = Sin(πx)/(πx) relation. This simple dataset is often used as a benchmark in machine learning studies since linear regression of a nonlinear function is impossible without a nonlinear transformation [20,21]. Each input value (x) was uniquely coded as a 2D pattern which was recorded on the SLM (see Supplementary Discussion 4 for details). By recording the nonlinearly propagated beam profile of many such input values, a linear regression method was performed on the output data (see Fig. 2 a). To measure the effectiveness of the spatiotemporal nonlinear propagation and assess the importance of the nonlinearity, we experimented with different pulse peak powers to control the level of nonlinearity. For low peak power (~1.14 kW), the nonlinearity is relatively weak, and the transmission through the fiber is very nearly a linear transformation (except for the square-law at the detector) and as a result, the performance is poor since the mapping is not linearly separable. Increasing the laser peak power results in nonlinear propagation, and we reached the best performance at around 3.43 kW laser peak power. For this optimum power level, correct outputs were estimated from unseen test inputs with a root-mean-squared error (RMSE) of 0.0671. Further power escalation gradually deteriorated the performance because it drives nonlinear pulse propagation to the Raman beam clean-up regime [34]. In this regime the projected beam profiles become virtually unaffected by the input data. In the other experiments reported below, the optical peak power used was 3.43 kW corresponding to the optimal peak power of the Sinc and COVID-19 diagnosis experiments.

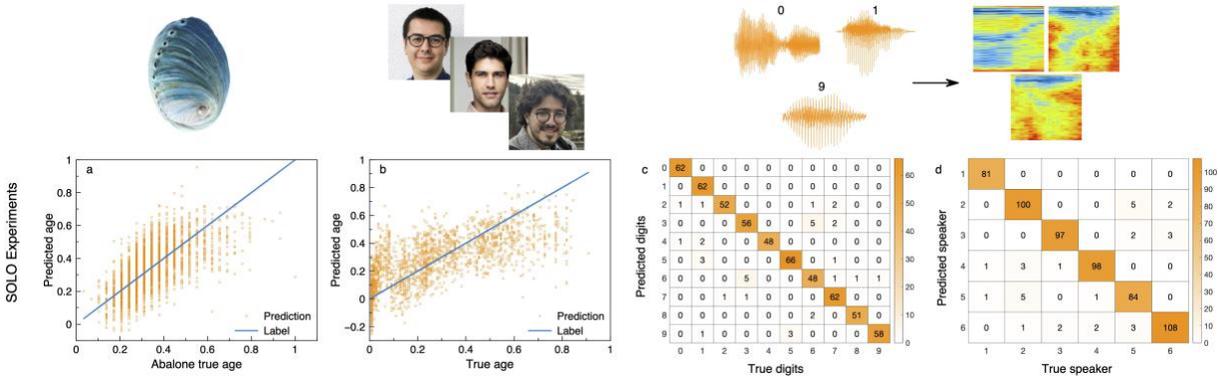

**Figure 3 Experimental (a-d) learning results for regression and classification tasks.** (a) Learning of abalone (multivariable) dataset. (b) Age prediction from face image dataset. (c) Confusion matrix for audio dataset digit classification. (d) Confusion matrix for audio dataset speaker classification.

*Abalone dataset*

The Sinc dataset demonstrates the interpolation capability of our optical computing framework; however, interpolation is not an adequate property for complex inference problems. Therefore, we moved to multivariable inference problems and we tested our computing method on the abalone dataset [35]. The abalone dataset consists of various physical features of sea snails in the dataset that are related to age (e.g., number of rings) that can be used for the prediction of the age of sea snails from eight different parameters. We recorded these 8 parameters on the SLM as a 4x2 matrix with proper pixel scaling. Similar to the Sinc function experiments, the recorded spatial distribution at the distal fiber facet was recorded, flattened (written as a long 1D vector) and fed to the decision layer to perform linear regression. (see Methods). Fig. 3 presents the true ages (Label) and the corresponding predictions; the figure indicates that the framework learns the ages of the abalone from spatially distributed independent variables with remarkable accuracy (RMSE of 0.126) compared with the output that takes normalized values between 0 and 1.

*Face image dataset*

Next, we tackled the problem of estimating the age of a person from an image of the person's face. A dataset containing 9780 images of faces of people from different gender and ethnicity with a long age span (0-116) is used [36]. The age is first normalized from 0 to 1. The number 1 represents the oldest person (116 years old). Here again, a single neuron is employed as the decision layer using the recorded fiber output intensity profiles. The achieved RMSE for age prediction is 0.167 normalized years. For the first 1000 samples the true ages (Labels) and predictions are shown in Fig. 3 b. Some predictions have negative values, which is impossible; however, this error is due to the final regression layer. This task is promising since image problems are massive and power-hungry in digital machine learning tools and gave birth to the convolutional neural network (CNN) architectures.

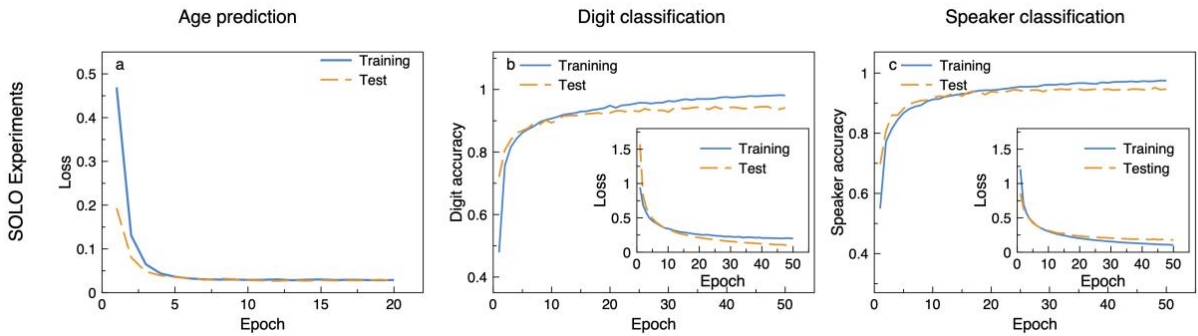

**Figure 4 Training of the experimental (a-c) results.** (a) Evolution of loss function (mean squared error) for age prediction dataset. (b) Evolution of accuracy and loss function (categorical cross entropy) for digit classification with audio dataset. (c) Evolution of accuracy and loss function (categorical cross entropy) for speaker classification with audio dataset.

*Audio digit dataset*

Classification of isolated audio records is one of the popular implementations of machine learning, which has a wide range of applications. We employed spoken digit classification to challenge the SOLO system. The audio digit classification dataset incorporates recordings of English digits by six distinct people [37]. Audio recordings are inherently time-varying signals. Following the standard approach, one-dimensional audio signals were converted to two-dimensional representation by generating so-called Mel spectrograms. These spectrograms of audio recordings were provided as inputs to the SLM. Similar to the previous dataset, the spectrograms are encoded on the pulses with high peak power. The output decision layer classifies the recorded respective fiber output intensity images and 94.5% accuracy over test data is obtained (see Fig. 3 c) for digit categorization task with frequency resolved beam profile measurement technique (see Methods).

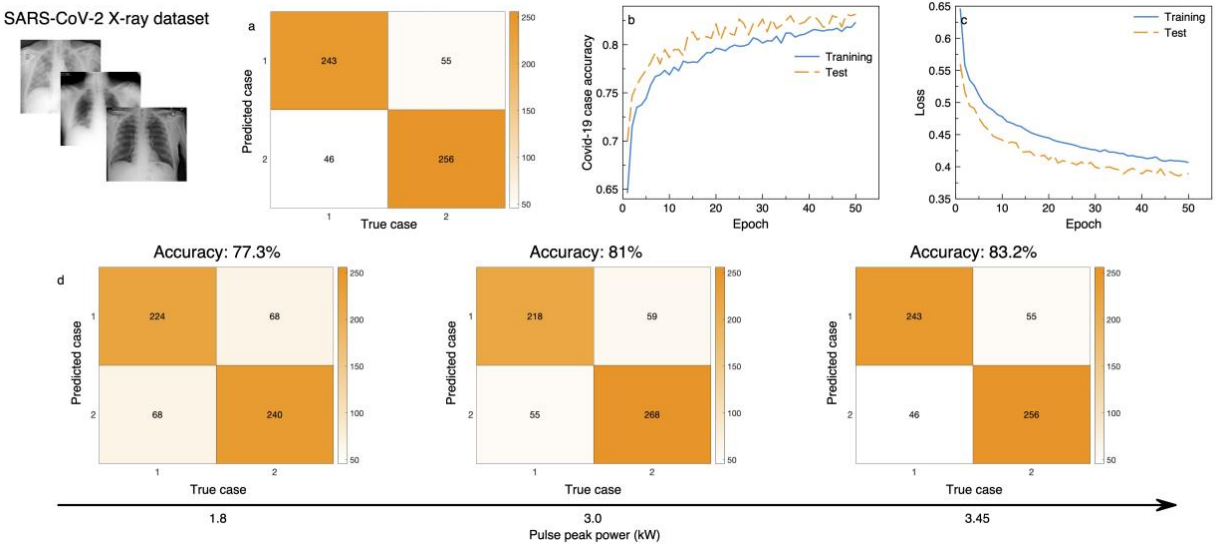

**Figure 5 Experimentally tested COVID-19 X-ray dataset.** (a) Confusion matrix for condition classification. Evolution of accuracy (b) and loss (c) functions (binary cross entropy) for COVID-19 X-ray dataset. (d) Measured effect of nonlinear pulse propagation on categorization accuracy.

To demonstrate the versatility of SOLO, we changed the task for the same dataset and aimed to differentiate the speaker from the audio record. Since the nonlinear transformation is independent of the task, we only updated the decision layer in SOLO and achieved 95.2% accuracy on test data as presented in Fig. 3 d with frequency resolved beam profile measurement technique (see Methods). The evolution of loss and accuracy (if applicable) functions for our digital decision layer with fiber simulation results are presented in Fig. 4 a-c.

*COVID-19 dataset*

Encouraged by the performance we obtained with the relatively simple tasks described so far, we tested SOLO with a difficult challenge of current interest by studying COVID-19 diagnosis with a dataset consisting of 3000 X-ray samples [38]. Similar to the audio dataset, the X-ray samples are applied to pulses as phase modulation and the corresponding fiber output intensity patterns were recorded. By performing classification in the decision layer, 83.2% accuracy over the unseen test set is achieved (see Fig. 5) with frequency resolved beam profile measurement technique (see Methods).

**Physical model**

The nonlinear mapping performed by the MMF can be investigated by the beam propagation method involving the fiber mode amplitudes (Eq. 1) [39]. In an ideal fiber without imperfections and bending, with low power pulse or continuous-wave light, only the phases of the mode coefficients change at different rates, due to modal and chromatic dispersion, without any intermodal power exchange. This behavior is captured by the first term in Eq. 1. This results in a linear transformation of the field as it propagates through the fiber.

Mode-coupling caused by perturbations due to fiber bending or by impurities, shown by matrix C, also acts as linear mixing (the second term in Eq. 1) [39, 40].

$$\frac{\partial A_p}{\partial z} = \underbrace{i\delta\beta_0^p A_p - \delta\beta_1^p \frac{\partial A_p}{\partial t} - i\frac{\beta_2}{2}\frac{\partial^2 A_p}{\partial t^2}}_{Dispersion} + \underbrace{i\sum_n C_{p,n} A_n}_{Linear\ Mode\ Coupling} + \underbrace{i\gamma \sum_{l,m,n} \eta_{p,l,m,n} A_l A_m A_n^*}_{Nonlinear\ Mode\ Coupling}$$

Equation 1

If the peak power of the pulse is high enough to induce nonlinear behavior in the material, nonlinear mode coupling takes place, and it results in a nonlinear operation on the information spatially encoded in the intense pulse throughout the fiber (the third term in Eq. 1). For each propagation step, the fiber modes couple to each other according to the linear coupling coefficients and the nonlinear coupling tensor, indicated by $\eta$. This nonlinear operator can be modeled at each propagation step by multiplying each three-element combination of mode coefficients with the related entry of the nonlinear mode coupling tensor (for details see Supplementary Discussion 1). In Eq.1, $\beta_2$ is the group velocity dispersion for the central frequency of the pulse and $\delta\beta_0^{(p)}$ ($\delta\beta_1^{(p)}$) is the difference between first (second) Taylor expansion coefficient of the propagation constant for corresponding and the fundamental mode.

**Numerical studies**

Our experiments demonstrated that the proposed optical computing framework has potential to learn with adequate performance. To understand the role of the nonlinearity in the MMF, and analyze its effect on learning, we performed time-dependent beam propagation method (TBPM) simulations (see Method section and Supplementary Discussion 2). Numerically studying pulse propagation in the 5m GRIN MMF for a dataset with 3000 samples requires approximately 2 years with our GPU parallelized simulation as explained in the Methods section. To reduce the computation time but include the required optical nonlinearity, we performed a rescaling of the propagation length and pulse peak power, also explained in the Methods section. We numerically studied the learning sinc function (see Supplementary Discussion 4), abalone dataset, face image dataset and audio dataset (see Supplementary Discussion 5). Note that the numerical simulation is only partial due to computational limitation and scalar. Due to the scalar nature of the simulation, the simulated fiber supports 120 spatial modes as opposed to the experiments where the test fiber supports 240 spatial modes. This difference causes information loss for the nonlinear mapping in the numerical studies. As a result, the data is not fully linearly separable after the numerically simulated operation whereas experimentally the data becomes linearly separable as evidenced by the flattening of the learning curves in Fig. 4.

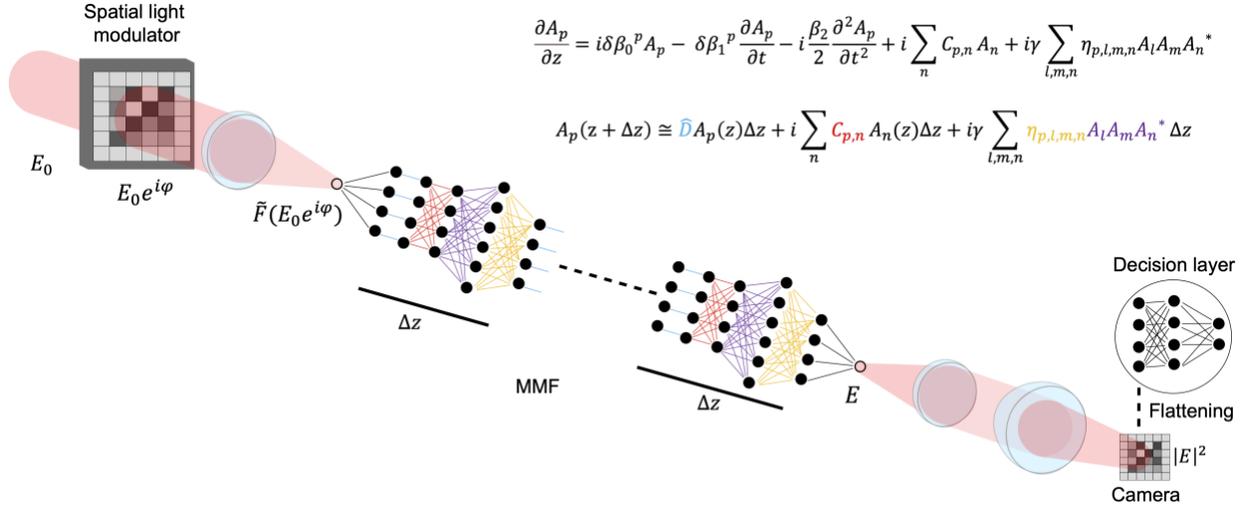

**Figure 6 Illustration of the spatiotemporal nonlinear pulse propagation as deep neural network architecture in the experimental setup**. Each elementary step Δz in the MMF is modeled as a cascade of linear and nonlinear network-like operators capturing the phase delay (in blue), the linear mode coupling due to bending and imperfections (in red), and the nonlinear mode coupling (in magenta and yellow).

We also simulated nonlinear beam propagation in GRIN MMF by encoding the COVID-19 dataset onto the optical pulses. 3000 X-ray samples are propagated numerically followed by classification of the resulting spatial distribution of the pulses. 70.8% accuracy is achieved (see Fig. 7 a-c), which is significantly lower than the experimentally obtained classification accuracy (83.2%). For categorization tasks, our numerical results offered lower performances than our experimental studies. These tasks require optically processing 2D information and the simulations were performed with a shorter fiber lengths and higher peak power due to the previously mentioned computational complexities. This simplification may not have captured the complex nonlinear mapping occurring in a longer fiber and lower peak power.

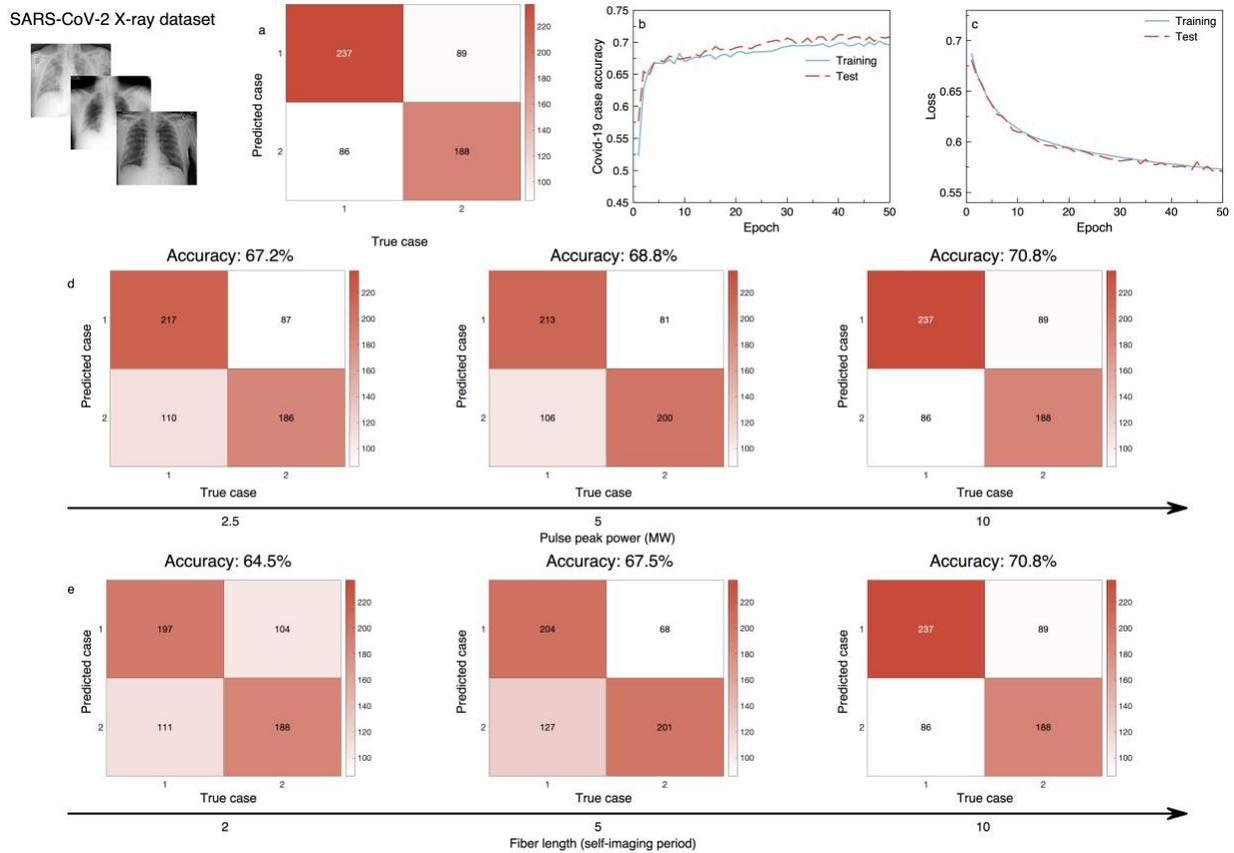

**Figure 7 Numerically tested COVID-19 X-ray dataset.** (a) Confusion matrix for condition classification. Evolution of accuracy (b) and loss (c) functions for COVID-19 X-ray dataset. Simulated effect of pulse peak power (d) and fiber length (e) on categorization accuracy.

To understand the impact of peak power and fiber length, we simulated the COVID-19 dataset with lower peak power levels and shorter fiber lengths. The numerically obtained 70.8% COVID-19 diagnosis accuracy decreased to 68.8% and 67.2% when the peak power respectively decreased to half and quarter of the initial power. Similar results were also obtained for shorter propagation lengths such as by decreasing the fiber length from 10 to 5 self-imaging period, we could achieve 67.5% diagnosis accuracy. A further decrease of fiber length by 2 self-imaging periods resulted in 64.5% diagnosis accuracy in our simulations. The numerically achieved confusion matrices for these studies are shown in Fig. 7. This simulation confirms the importance of high intensity light in learning ability.

## Discussion

The present study reveals that the nonlinear interactions in spatiotemporal pulse propagation in MMFs is a key element for learning. It is also important to understand how the performance of SOLO scales with the input data size. We first analyze how the power scales with the input size for SOLO. The number of modes N in a MMF scales proportionally to the fiber core area, hence the optical power necessary to maintain the same intensity scales linearly with the number of modes. Since a fiber having N modes can accommodate an N-dimensional input (law of etendue), the optical power scales with the size of the dimensional input N. The GRIN MMF used in the present study supports 240 modes (counting the polarization degeneracy). In the experiments, learning reached an optimum for a pulse peak power of 3.4 kW for nonlinear optical effects which corresponds to 4.4 mW average optical power for 125 kHz repetition rate and 10 ps pulse duration. Thus, to perform the computation in our experiments, the required average optical power is 18 µW per fiber mode (4.4 mW/240). Here we would like to remark that the reported required average optical power per fiber mode is depending on the repetition rate of the laser. Since the speed of the SLM and camera is relatively low, one can decrease the repetition rate of the laser source to match the SLM. Thus, while keeping the peak power and the accumulated optical nonlinearity at the same level, less power per mode requirements for computation can be achieved.

In terms of optical computing operations, we can assume that the number of operations is at least of the order of $N^2$, since the input of size N is first multiplied with the number of modes (mode decomposition) and then each of the N modes is operated upon. This can be seen from the propagation Eq. 1 or its implementation as a network in Fig. 6. The nonlinear coupling tensor $\eta$ has the largest terms for self-phase and cross phase modulation [39]. Even keeping only these terms, the number of multiplications reduces to $N^2$. As Fig. 6 suggest, this latter computation is performed many times in the fiber.

With current SLM technology, the number of inputs N can reach up to $10^7$ pixels with a 60 Hz refresh rate [41]. Increasing the number of fiber modes to $10^7$ can be done by employing large core MMFs and/or multicore MMFs. The number of Operations per seconds would then be $N^2$ x SLM refresh rate, or more than 6 PetaOps/s. The digital single layer network N x 1 following the optical computation would require only $6 \bullet 10^7$ FLOPs.

The digital counterparts of the SOLO can be categorized into three categories. The first approach can be the simulations with exact experimental parameters. As we already saw such an approach will require more than 2 years of GPU computation for a dataset with 3000 samples when the simulation grid sizes are set to the parameters explained in the Methods section of this paper. The second approach can be to compare SOLO to a deep neural network that is trained to learn the spatiotemporal pulse propagation in MMFs. This approach has been achieved for linear propagation in MMFs with particular datasets [42,43] with networks having at least >14 layers and >50 million parameters (weights). Nonlinear propagation studies are yet to be investigated but it is reasonable to expect that more complex networks would be required to learn nonlinear propagation in MMF. The last approach can be the standard deep neural networks whose structure is unrelated to SOLO for each specific task. This approach requires designing specific network architectures for each dataset type.

SOLO is a computing framework on which different problems were embedded and were experimentally implemented. Our study demonstrated calculations that yield performance comparable to convolutional neural networks and higher than fully connected ELM architectures (see Supplementary Discussion 12). With digital or optical feedback, the present optical paradigm could act as a reservoir network structure [44]. Our tests targeted supervised learning examples, yet unsupervised learning is possible with the proposed technique due to the label-free nonlinear projection behavior of the spatiotemporal pulse propagation. To increase the performance of SOLO or to adapt it for more challenging tasks, the decision layer used in the present studies can be modified with additional layers. Thus, SOLO can compute information as a fast and efficient front-end module.

To evaluate the robustness of the present optical computing method, we repeated the experiments for the COVID-19 dataset a weeks after the prior measurements presented in Fig. 5. Without requiring a calibration, we obtained the similar learning performance level, around 82% accuracy on the test set, for diagnosing COVID-19 from X-ray images (see Supplementary Discussion 8). Furthermore, we performed detailed analysis to determine the stability of the setup. As an analog system, equipment (Laser source, SLM) used in the SOLO experiments inherit noise-like behaviors such as laser pointing stability, SLM stability. In our tests (see Supplementary Discussion 10), we obtained high stability with 12.63 signal to noise ratio.

## Outlook and Conclusion

The presented optical computing framework can be further improved with an active MMF scheme where the fiber is mechanically perturbed [45] or the pump light is also shaped to control spatiotemporal nonlinear propagation. Different cases of adaptive pumping in fiber amplifiers are already demonstrated in the literature [46,47]. Such an implementation may lead to optically controllable computing with nonlinear fiber optics.

We envision that another implementation of SOLO can be realized with silicon-on-Insulator technology. This technology enables optical functions on integrated circuits, which already resulted in many useful applications [48]. Nonlinear silicon photonics already demonstrated supercontinuum generation through self-phase modulation, light amplification using the Raman effect and matrix convolution operations [49,50]. By leveraging the existing integrated silicon photonic manufacturing platform, it is possible to implement the machine learning that we demonstrated in optical fibers.

In conclusion, we have shown how spatiotemporal nonlinear pulse propagation in MMFs can optically process information to compute complex machine learning tasks that only sophisticated deep neural networks can tackle. In our benchmarks, the proposed optical computing platform performs as powerful as its digital counterparts for different tasks. With better energy efficiency that previous proposals and a path to PetaOPs scalability, SOLO provides a novel path toward supercomputer-level optical computation.

## Method

*Experimental setup*

As the light source, an Yb fiber laser (Amplitude Laser-Satsuma) that produces 10ps pulses with a 125 kHz repetition rate is selected. The pulse is centered around 1033 nm with a width of 10 nm. The linearly polarized Gaussian laser output beam is shaped via a phase-only SLM (Holoeye Pluto-NIRII), an 8um pixel pitch and 60 Hz speed. The SLM prints the desired input pattern on top of a grating phase pattern that expels unmodulated light. We used 5m of a commercial GRIN 50/125 multimode fiber with NA of 0.2; this fiber allows 120 modes per polarization for the given excitation. The phase-modulated light from SLM is imaged onto the MMF with a 15mm lens focal length. The information beam covers the whole MMF core area. The beam-core overlap is checked by imaging the back reflection of the proximal fiber side (not shown in Fig 1). The distal fiber side is magnified 12.5 times through a 4f imaging system and recorded by a camera with a 5.2um pixel pitch. As an alternative method, instead of 4f imaging, frequency-resolved spatial measurement with a dispersive optical element (grating with a 600line/mm period) is used as presented in Fig 1. For categorization tasks (audio digit and Covid-19 datasets) significant performance increases are observed and reported here. We monitored fiber output power after and before the MMF continuously. Various neutral density filters are embedded to avoid camera saturation. The pulse power and width are optimized so that the pulse conserves its temporal unity (no temporal splitting) and maximizes spatial interactions.

*Numerical Simulations*

We implemented a GPU parallelized time-dependent beam propagation method (TBPM) in Python to simulate sufficiently fast nonlinear pulse propagation in the fiber. TBPM simulations often require long computational times due to heavy multidimensional fast Fourier transform calculations. The launched pulses centered at 1030 nm with one ps duration were numerically propagated for 10 self-imaging periods distance. In the experiment, the fiber length is 5 m. To reach a manageable computing time for the datasets with 3000 samples, we performed a rescaling of the propagation length from 5 m to ~5.5 mm. To generate significantly nonlinear spatiotemporal evolution in such a short propagation, we increased the pulse peak power to 10 MW. The time window of simulation is 20ps with 9.8 fs resolution and the spatial window is set as a 64x64 spatial grid. To properly simulate the graded-index MMF's spatial self-imaging, the numerical integration step is set to sample each self-imaging period 16 times. To create an absorptive boundary condition around the core we truncated the parabolic fiber index profile with the super-Gaussian filter. We matched the launched Gaussian beam diameter ($1/e^2$) to fiber core size (50 μm). For our studies, we encoded data into the beam as a multiplied phase information. After propagation, the obtained pulse is time-averaged and converted into normalized intensity images. There are several ways of converting images into one-dimensional representations. For simplicity, we used a flattened version of downsampled images as an output vector. Finally, flattened output vectors are linearly fitted using the standard Linear Regression method.

# Data and code availability

The numerical data used in this work and a public version of the codes are available at https://github.com/ugurtegin/Nonlinear_MMF_Network.

# Acknowledgements

The authors thank Niyazi Ulaş Dinç and Prof. Yaser Abu-Mostafa for fruitful discussions.

## Author contributions

U.T., M.Y. and I.O. performed simulations and experiments, C.M and D.P. supervised and directed the project. All the authors participated in the analysis of the data and the writing process of the manuscript.

*Supplementary Material of*
Scalable Optical Learning Operator

Uğur Teğin[1,2,*], Mustafa Yıldırım[2], İlker Oğuz[1,2], Christophe Moser[2] and Demetri Psaltis[1]

[1] Optics Laboratory, École Polytechnique Fédérale de Lausanne, Switzerland
[2] Laboratory of Applied Photonics Devices, École Polytechnique Fédérale de Lausanne, Switzerland
* ugur.tegin@epfl.ch

**Supplementary Discussion 1: Physical Model of the Computation Framework**

The optical beam at any position of the optical fiber can be decomposed into spatial modes of the fiber. In Eq. 1, $E(\rho, \varphi, \omega)$ is the electric field of the light, $A_l$ is the envelope of the corresponding mode along the propagation direction $z$ and $F_l$ is the mode shape. Eq.2 shows the solution of the mode shape $F_l$ for graded index fibers having relative index difference of $\Delta$ and radius of R, $L_p$ is the generalized Laguerre polynomial [S1,S2]. The modes propagation constant $\beta_{p,m}$ are calculated using Eq. 3.

$$E(\rho, \varphi, \omega) = \sum_l F_l(\rho, \varphi, \omega) \, e^{i\beta_l(\omega)z} \, \tilde{A}_l(z, \omega)$$

(Supp.Eq. 1)

$$F_{p,m}(\rho, \varphi, \omega) = \sqrt{\frac{p!}{\pi(p+|m|)!}} \frac{\rho^{|m|}}{(w_0/\sqrt{2})^{|m|+1}} \, e^{-\rho/w_0^2} \, L_p^{|m|}\left(2\frac{\rho^2}{w_0}\right) e^{im\varphi}$$

(Supp.Eq. 2)

$$\beta_{p,m}(\omega) = k\sqrt{1 - \frac{2\sqrt{\Delta}}{kR}(2p + |m| + 1)}$$

(Supp.Eq. 3)

Propagation of an intense pulse inside an optical fiber can be analyzed following this representation. Eq. 4 represents the nonlinear spatiotemporal evolution of each mode. Each mode is coupled to the others through the nonlinearity tensor coefficient $\eta_{p,l,m,n}$ which models nonlinear intermodal and intramodal effects, and through a linear coupling coefficient ($C_{p,n}$) which expresses mode coupling due to perturbations to the ideal fiber shape and refractive index distribution, such as bending and impurities. The linear coupling coefficient ($C_{p,n}$) relates perturbation in permittivity (or refractive index) to intermodal model coupling by calculating the overlap integral with the corresponding mode shapes [S3]. Similarly, the nonlinearity tensor coefficient ($\eta_{p,l,m,n}$) is computed with the normalized overlap integral of modes. We computed the nonlinear coupling tensor between the modes for our GRIN-50/125 fiber. The tensor has a $120^4$ size and computing all nonlinear terms took two and a half months on a server computer with 2x Intel Xeon CPU E5-2670 and 384 GB of RAM. The cross-phase modulation coefficients ($\eta_{p,p,q,q}$, inter modal nonlinearities) are shown in Fig. 1, where the diagonal terms correspond to self-phase modulation ($\eta_{p,p,p,p}$, intramodal nonlinearities). This demonstrates the richness of the nonlinear interaction that SOLO relies upon. Note that four-wave mixing could not be shown on the graph due to its dimensionality.

$$\frac{\partial A_p}{\partial z} = \underbrace{i\delta\beta_0{}^p A_p - \delta\beta_1{}^p \frac{\partial A_p}{\partial t} - i\frac{\beta_2}{2}\frac{\partial^2 A_p}{\partial t^2}}_{Propagation\ and\ Dispersion} + \underbrace{i\sum_n C_{p,n} A_n}_{Linear\ Mode\ Coupling} + \underbrace{i\gamma \sum_{l,m,n} \eta_{p,l,m,n} A_l A_m A_n{}^*}_{Nonlinear\ Mode\ Coupling}$$

(Supp.Eq. 4)

$$\eta_{p,l,m,n} = \frac{\iint dx\,dy\, F_p F_l F_m F_n}{[\iint dx\,dy\, F_p \iint dx\,dy\, F_l \iint dx\,dy\, F_m \iint dx\,dy\, F_n]^2}$$

(Supp.Eq. 5)

$$C_{p,n} = \frac{\omega}{2} \iint dx\,dy\, (\widetilde{\epsilon^*} - \epsilon) F_p F_n^*$$

(Supp.Eq. 6)

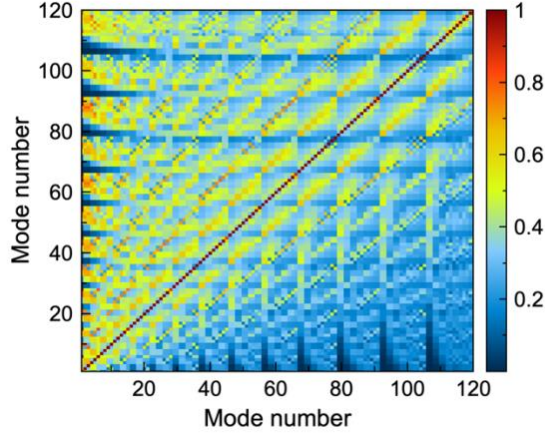

**Supplementary Figure 1:** Intra- and Intermodal nonlinear coupling coefficients ($\eta_{p,p,q,q}, \eta_{p,p,p,p}$)

**Supplementary Discussion 2: Time Dependent Beam Propagation**

Instead of considering individual mode field propagation, we use. Eq.7 to describe the total field propagation in a GRIN fiber without perturbations. We numerically implemented Eq.7 using symmetrized split-step Fourier Method. The codes are implemented in Python + Cupy library which made it possible to utilize powerful GPUs. The time steps are selected at the Nyquist rate defined by the wavelengths range given in Eq. 8-9. We used the self-imaging period as one of our control knobs in simulations using the equations in Supp. Eq.10 [S2]. $A$ in the Supp. Eq. 8 is the slowly varying envelope at the center frequency.

The datasets are divided with a ratio of 0.2 for training (2400 samples) and validation (600 samples). The regression is implemented using Scikit-learn or Tensorflow on Google Colab cloud service which provides an Intel Xeon CPU and Nvidia Tesla V100 GPU. We also used single dense layer without nonlinear activations for linear regression.

$$\frac{\partial A}{\partial z} = \frac{i}{2k_0}\left(\frac{\partial^2 A}{\partial x^2} + \frac{\partial^2 A}{\partial y^2}\right) - i\frac{\beta_2}{2}\frac{\partial^2 A}{\partial t^2} + \frac{\beta_3}{6}\frac{\partial^3 A}{\partial t^3} - \frac{ik_0\Delta(x^2+y^2)A}{R^2} + i\gamma|A|^2 A$$

(Supp.Eq. 7)

$$\lambda_{min} = \frac{1}{\frac{1}{2c\Delta t} + \frac{1}{\lambda_0}}$$

(Supp.Eq. 8)

$$\lambda_{max} = \cfrac{1}{-\cfrac{1}{2c\Delta t} + \cfrac{1}{\lambda_0}}$$

(Supp.Eq. 9)

$$Self\ imaging\ period = \frac{\pi R}{\sqrt{2\Delta}}$$

(Supp.Eq. 10)

**Supplementary Discussion 3: Data encoding to SLM**

Images are 2D arrays, therefore the image datasets are directly mapped to SLM pixels. We illuminated the 600-by-600 pixels central region of the SLM and all images are scaled to that size to cover the entire beam. A blazed grating is added to the pattern to prevent unmodulated DC light to enter the fiber. Encoding 2D arrays are relatively easy than a scalar or 1D input. To handle a scalar input (such as for Sinc experiment), we mapped the scalar value to a 2D array by multiplying the value through a fixed random 2D matrix. This provides unique 2D matrixes for every distinct input value. For a 1D input, we simply converted them to 2D and upscaled to the illumination pixel range.

**Supplementary Discussion 4: Single value regression of Sinc function**

First, we tested the nonlinear information transformation ability of spatiotemporal propagation of high peak power pulses numerically by performing time-dependent beam propagation method (TBPM) simulations (Supplementary Discussion 2). Our first numerical simulation was learning the Sinc function input-output relation numerically duplicating the experiment we described above. The Sinc is a simple nonlinear function that cannot be learned with a single-layer network and it has been used as a standard benchmark to validate learning methods [18,19]. We used 3000 randomly generated samples, which lie in [-π,π] to cover the Sinc function's characteristic behavior. We fed the generated samples to TBPM by expanding scalar values to two-dimensional form using a random mask and calculated the nonlinear pulse propagation in the GRIN MMF. The projected intensity distribution at the distal end of the fiber is considered as the nonlinearly transformed information. The linear regression parameters are retrieved from the training data, and the overall performance is assessed by the test data. A remarkable learning performance with 0.0039 root-mean-squared error (RMSE) for the test data is obtained. The achieved performance shows that spatiotemporal fiber nonlinearity provides a significant contribution to learning ability. The result proves that our computational device transformed the input space to a higher dimensional space efficiently such that the proposed framework interpolates a function for the unseen data.

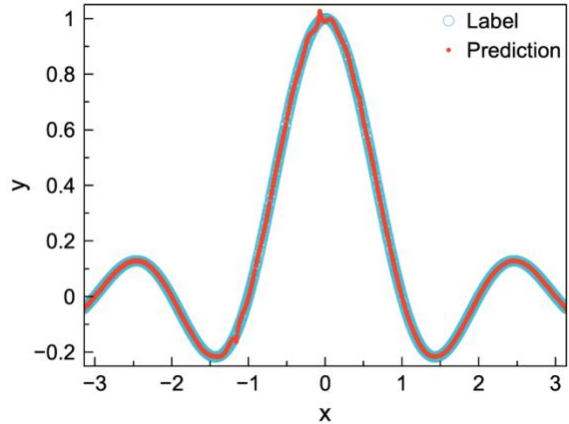

**Supplementary Figure 2:** Learning the Sinc function from simulated data

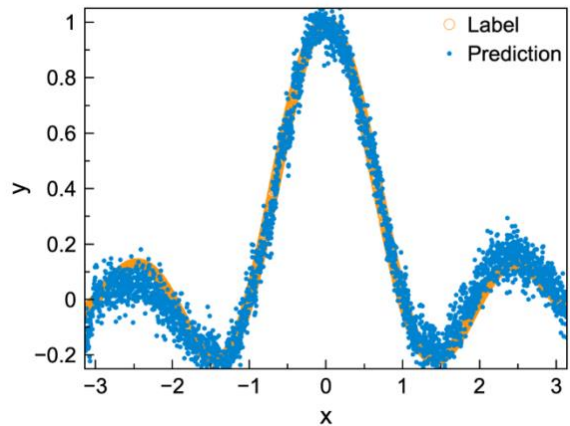

**Supplementary Figure 3:** Learning the Sinc function from experimental data

**Supplementary Discussion 5: Results of numerical studies**

We simulated the nonlinear pulse propagation with the abalone dataset information to perform multivariable regression. We encoded the abalone features as the spatial phase distribution of a pulse in our numerical implementation onto the input beam. Similar to the Sinc function, a decision layer to perform linear regression is employed and we obtained an age prediction with remarkable accuracy (RMSE of 0.0831). Supp. Fig. 4 a presents normalized correct ages and predictions. We continued our numerical studies with the face image dataset. By encoding different human face images into the simulated pulse, each person's age on the images was estimated, and an

RMSE of 0. 2175 on normalized output values indicated again close correspondence with the experimental studies (see Supp. Fig. 4 b).

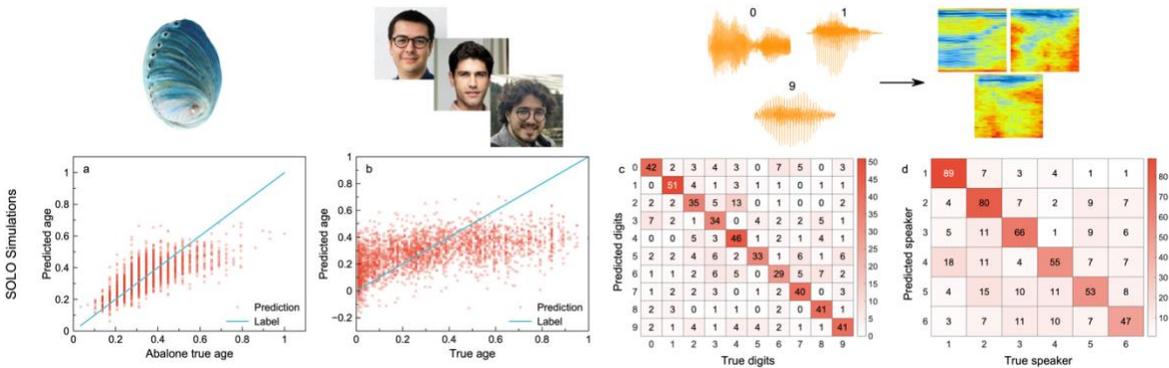

**Supplementary Figure 4: Experimental (a-d) learning results for regression and classification tasks.** (a) Learning of abalone (multivariable) dataset. (b) Age prediction from face image dataset. (c) Confusion matrix for audio dataset digit classification. (d) Confusion matrix for audio dataset speaker classification.

As indicated in our experimental studies, the audio data can also be converted to a two-dimensional format and regarded as an image analysis task by calculating the related spectrogram. This way, we simulated the nonlinear propagation of pulses for audio digit for categorization purposes. By taking the fiber output beam shapes as inputs to a single layer classifyer gave approximately 68% accuracy as shown in Supp. Fig. 4 c. Similar to our experimental investigations, we updated the decision layer and tried to differentiate the speaker from the audio record. In our numerical studies, we obtained 61% accuracy over the unseen test set (see Supp. Fig. 4 e). The evolution of loss and accuracy (if applicable) functions for our digital decision layer with experimental results are presented in Fig. 5 a-c.

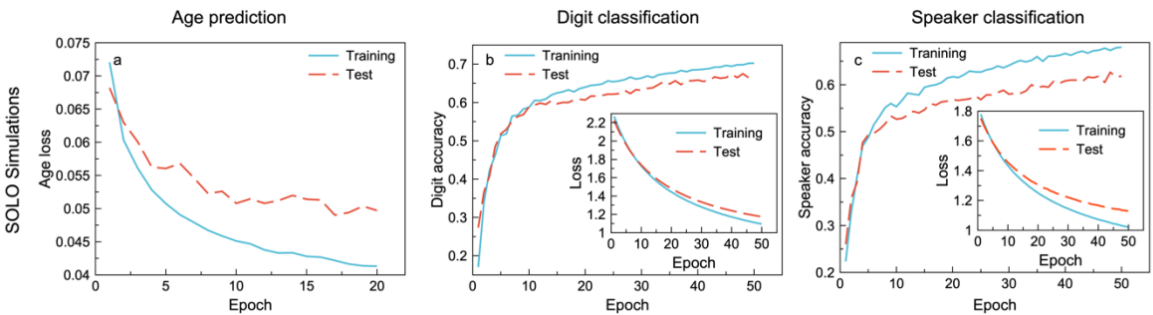

**Supplementary Figure 5: Training of the numerical (a-c) results**. (a) Evolution of loss function for age prediction dataset. (b) Evolution of accuracy and loss function for digit classification with audio dataset. (c) Evolution of accuracy and loss function for speaker classification with audio dataset.

**Supplementary Discussion 6: Power Oscillations during experiment**

Learning is related to the amount of nonlinearity (or power). Therefore, stable power is required during experiments. Supp. Fig. 6 shows power fluctuations recorded at the fiber end. There is a small variation due to the modulated diffraction efficiency of grating that depends on the encoded patterns. During this measurement, the mean output average power is 1.58 mW and standard deviation is 0.0292 which results in 1.84% power oscillation for the dataset.

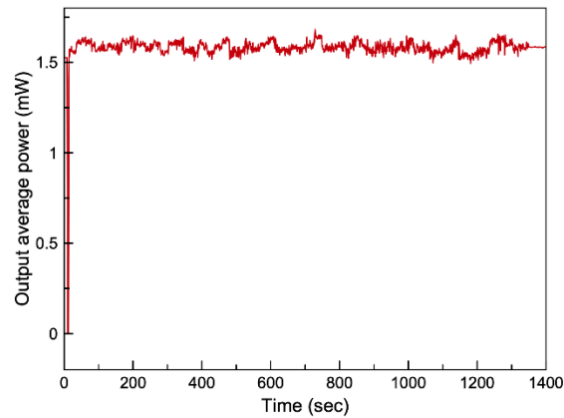

**Supplementary Figure 6:** Output power fluctuations of Face dataset

**Supplementary Discussion 7: Peak power vs Beam and Spectrum Analysis**

The effect of the nonlinearity on the output beam shape and spectrum are shown in Supp. Fig. 7. In these experiments, a symmetric Gaussian beam with flat phase is injected without any encoding. The power of the beam at the output of the fiber is concentrated in the center. In fig. 7, we observe that an increase of peak power spreads the spectrum and the spatial distribution of the output beam initially spread from center (Supp. Fig. 7 a-b). Raman beam cleaning [S4] then becomes dominant when the power further increase, which creates again a beam with power concentrated in the center (Supp. Fig. 11).

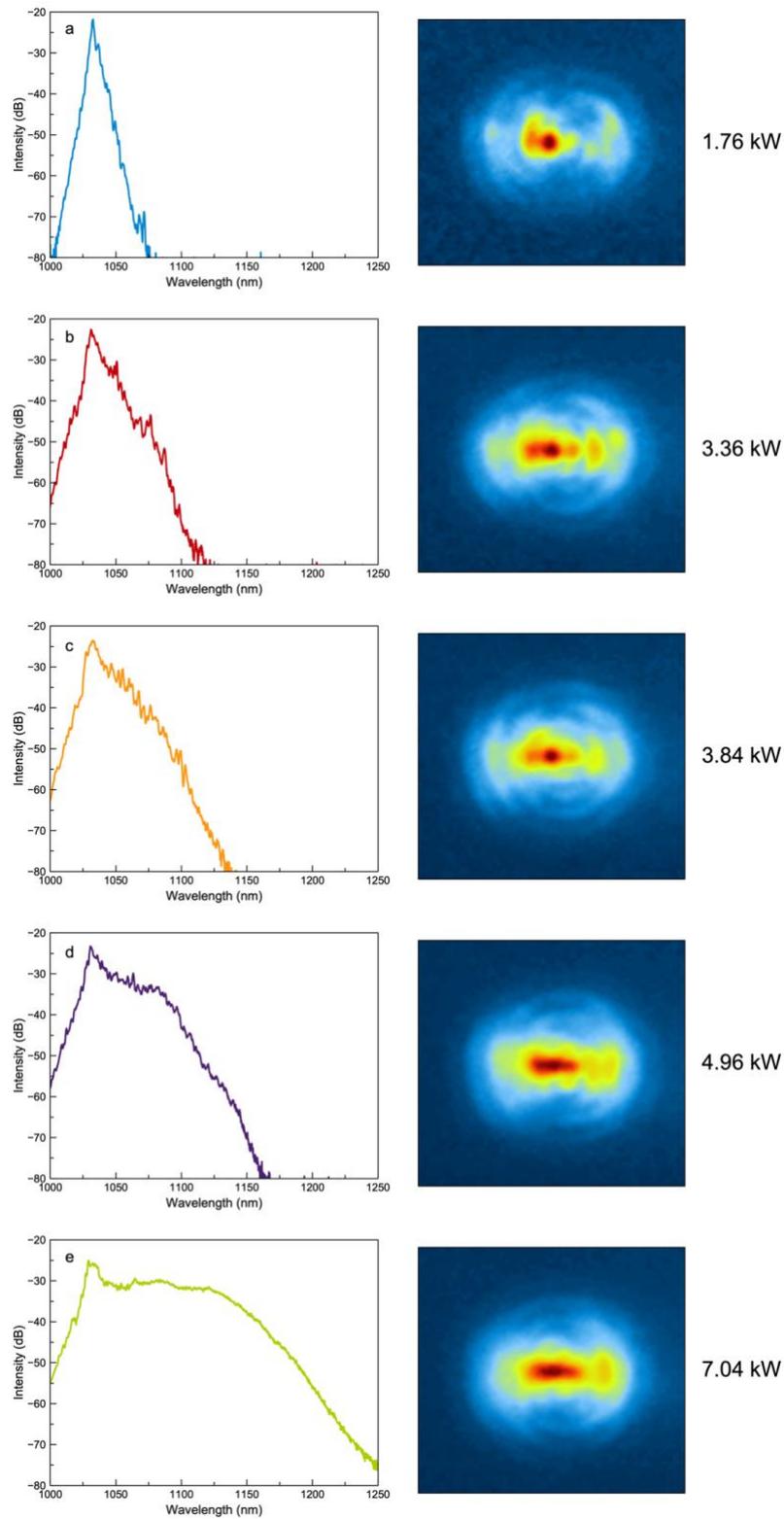

**Supplementary Figure 7:** Spectrum and output beam evolution at the end of the test fiber with increasing pulse peak power.

**Supplementary Discussion 8: Robustness of experiments**

The robustness test results of the experimentally demonstrated optical computing method are presented in this section. The exact same experiments with the COVID 19 X-ray images were performed again at a week interval. Around 82% accuracy over test data is achieved like the results obtained in Fig.5 in the main text. These results validate the robustness of the present optical computing framework.

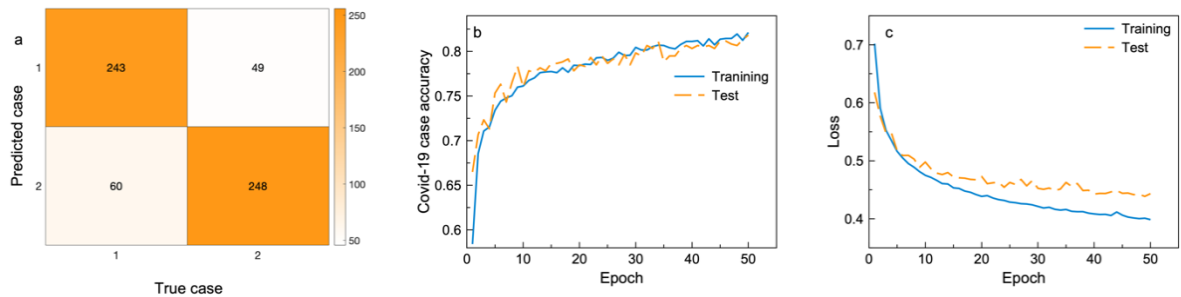

**Supplementary Figure 8: Experimentally tested COVID-19 X-raydataset.** (a) Confusion matrix for condition classification. Evolution of accuracy (b) and loss (c) functions for COVID-19 X-ray dataset.

**Supplementary Discussion 9: Covid-19 X-ray dataset outputs**

An example of input and output relation in SOLO is presented in Supp. Fig. 9. Here the input images are loaded on the SLM. After propagation through the fiber, the beam profile is recorded corresponding to each respective input. These beam profiles contain the nonlinearly processed input information for the decision layer, as explained in detail in the Methods section.

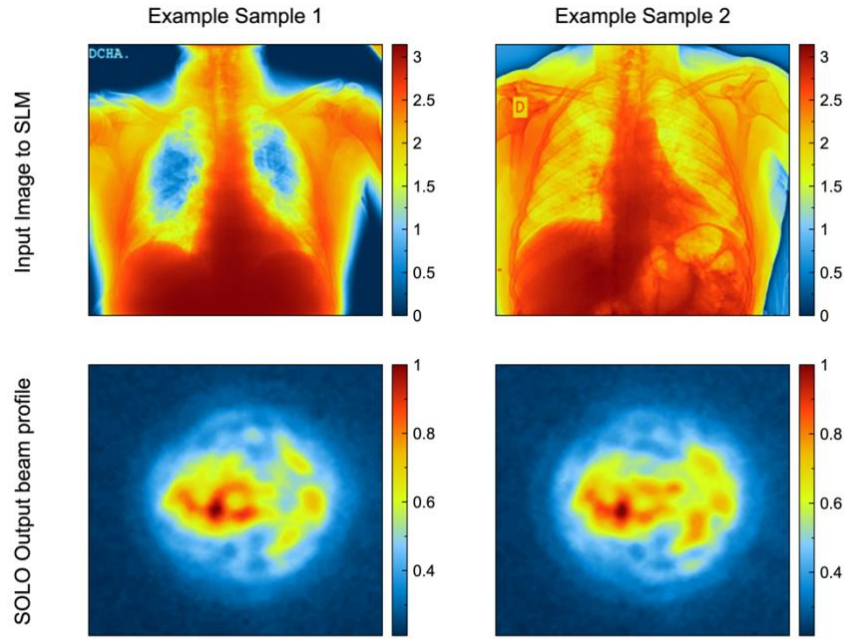

**Supplementary Figure 9:** Experimentally used input images and SOLO output beam profiles for COVID-19 X-ray dataset.

**Supplementary Discussion 10: Stability and noise of experiments**

To measure the stability and the noise of the experiments, we send the same data (1st image in Covid-19 X-ray dataset) for an hour in the SOLO experiment. In every step, the image is encoded to SLM from scratch and the fiber output measurements are performed. The Supp. Fig. 10. and Supp. Fig. 11. shows the changes in the obtained fiber output images. The measurement time for the dataset of a 3000 sample in our experiments is 30 minutes. For this time interval, the average RMSE in the presented measurement in Supplementary Figure 11 is 0.079 which corresponds to a 12.63 signal to noise ratio (SNR). With the log2(SNR) relation, the bit accuracy of the experiment is around 3.65. As another measure of stability, after normalizing each image to unit power, average of standard deviation of all pixel values across frames is calculated to be 4.62% for 30 minutes and 5.48% for 60 minutes of data acquisition.

The obtained SNR and RMSE values include all the possible noise and drift sources that can take place in our experimental configuration. These possible noise and drift sources include laser pointing stability, SLM's electronic and temperature noise and optical alignment changes.

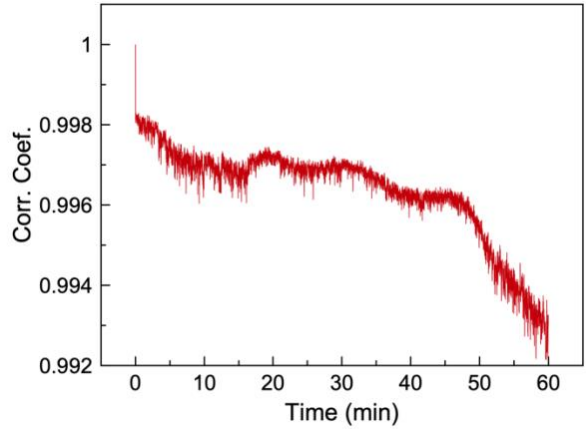

**Supplementary Figure 10:** Pearson correlation coefficient between the first fiber output beam profile and the following measurements over an hour.

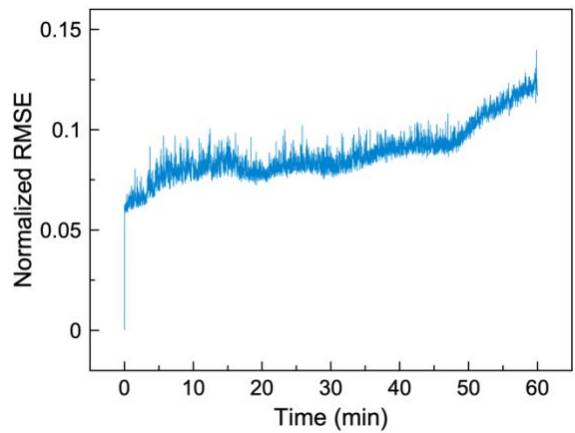

**Supplementary Figure 11:** Normalized root mean squared error between the first fiber output beam profile and the following measurements over an hour.

**Supplementary Discussion 11: Effect of peak power for other datasets**

Similar to the sinc function dataset presented in Fig. 2, peak power is the only controlling factor for adjusting the accumulated nonlinear interactions inside the fiber for the reported configuration. As an example, impact of the peak power on the accuracy is demonstrated in Supplementary Table 1 for the audio digit and Covid-19 X-ray datasets.

| Peak power (kW) | Covid-19 dataset acc. | Audio digit dataset acc. |
|:---:|:---:|:---:|
| 1.76 | 77% | 91% |
| 3.11 | 78% | 92% |
| 3.47 | 83% | 95% |
| 3.77 | 78% | 93% |
| 7.09 | 77% | 91% |

**Supplementary Table 1**

**Supplementary Discussion 12: Comparison of SOLO with digital neural networks**

To make a comparison of the performance of SOLO with its digital counterparts, we performed the classification tasks with Covid-19 and Audio dataset by using an ELM and convolutional neural network (CNN), similar to LeNet. The ELM is consisting of 1 layer with fixed weights and nonlinear (sigmoid) activation function and a following trainable layer. The CNN is consisting of 3 convolutional layers (with maxpool and dropout features) and 2 dense layers. The accuracy comparison between the computation architectures is presented in Supplementary Table 2.

| Architecture | Acc. for Covid-19 | Acc. for Audio Digit | Acc. for Audio Speaker |
|---|---|---|---|
| **SOLO** | 83% | 95% | 95% |
| **ELM** | 75% | 82% | 88% |
| **CNN (3C+2D)** | 91% | 98% | 99% |

**Supplementary Table 2**

# Supplementary References

# Authors' Note:

In addition to the other changes, in our revised arXiv preprint we updated experiments and results for Audio digit and Covid-19 datasets.

- We discovered that in our initial experiments with Audio digit and Covid-19 CT-scan datasets the measurements were performed in an ordered (unshuffled) manner. In such a scenario, any experimental drift can act as a performance-increasing effect on the score of the SOLO. Thus, we repeated our experiments with Audio digit and Covid-19 datasets in a shuffled order and updated our results in this revised manuscript.

- We introduced a secondary data collection method, in addition to the 4F imaging method, inspired by frequency-resolved measurements with a dispersive optical element (diffraction grating) as it is shown in the updated Fig.1 in our revised manuscript. In our experiments, we experienced that frequency-resolved measurement provides higher accuracy and better learning curves for classification tasks.

- We replaced the Covid-19 CT-scan dataset we used in our initial manuscript with a Covid-19 X-ray dataset. The CT-scan dataset contains 2482 scans acquired from 120 patients. This results in repetitive CT-scan images and intrinsic clustering among the data points. Thus, we changed the dataset with an X-ray image dataset for our Covid-19 classification task. In this new dataset which contains 3000 images, each X-ray image is associated with a different individual thus it does not contain repetitive samples as is the case for the Covid-19 CT-scan dataset. While updating our Covid-19 prediction results in our manuscript, we also compared our performance with a digital convolutional neural network and reported the performance table in our supplementary document.